\documentclass[journal]{IEEEtran}

\ifCLASSOPTIONcompsoc
  \usepackage[nocompress]{cite}
\else
  \usepackage{cite}
\fi
\ifCLASSINFOpdf

\else

\fi

\usepackage[cmex10]{amsmath}
\usepackage{graphicx}

\usepackage{subcaption}

\usepackage{array}

\usepackage{booktabs}
\usepackage{mathtools}

\usepackage{algorithm}
\usepackage{algorithmic}
\makeatletter
\newcommand\fs@norules{\def\@fs@cfont{\bfseries}\let\@fs@capt\floatc@ruled
  \def\@fs@pre{}%
  \def\@fs@post{}%
  \def\@fs@mid{\kern3pt}%
  \let\@fs@iftopcapt\iftrue}
\makeatother
\floatstyle{norules}
\begin{document}

\title{Boosting the Performance of Content Centric Networking using Delay Tolerant Networking Mechanisms}
\author{\IEEEauthorblockN{Hasan M A Islam$^1$, Dimitris Chatzopoulos$^2$, Dmitrij Lagutin$^1$, Pan Hui$^2$ and Antti Yl{\"a}-J{\"a}{\"a}ski$^1$} \\
\IEEEauthorblockA{$^1$Aalto University, $^2$Hong Kong University of Science and Technology} 
\vspace{-0.3cm}
}

\maketitle

\begin{abstract}
Content-Centric Networking (CCN) introduces a paradigm shift from a host centric to an information centric communication model for Future Internet architectures. It supports the retrieval of a particular content regardless of the physical location of the content. Content caching and content delivery networks are the most popular approaches to deal with the inherent issues of content delivery on the Internet that are caused by its design. Moreover, intermittently connected mobile environments or disruptive networks present a significant challenge to CCN deployment. In this paper, we consider the possibility of using mobile users in improving the efficiency of content delivery. Mobile users are producing a significant fraction of the total internet traffic and modern mobile devices have enough storage to cache the downloaded content that may interest other mobile users for a short period too. We present an analytical model of the content centric networking framework that integrates a Delay Tolerant Networking (DTN) architecture into the native CCN, and we present large scale simulation results. Caching on mobile devices can improve the content retrieval time by more than 50$\%$, while the fraction of the requests that are delivered from other mobile devices can be more than 75$\%$ in many cases.  
\end{abstract}

\IEEEpeerreviewmaketitle

\section{Introduction}

Today's Internet architecture relies on the fundamental assumption that there exists an end-to-end path between the source and destination during the communication session. However, the vast majority of Internet usage is dominated by content distribution and retrieval involving a large amount of digital content and this makes the conventional Internet architecture inefficient. In response, Information-Centric Networking (ICN)~\cite{xylomenos2014survey} emerges as a paradigm shift from a host centric to an information centric communication model. It supports the retrieval of a particular content without any reference to the physical location of the content. \textit{Named data} is the central element of ICN communication instead of its physical location. When a node needs content, it sends a request for a particular content. If any node on the route of the request has the content in its content store, it replies with that content to the request. The main argument for this architectural shift is that named data provide better abstraction than named hosts. 

Among all the ICN proposals, Content Centric Networking (CCN) architecture \cite{jacobson2009networking} is gaining more and more interest for its architectural design. CCN supports two types of messages: \textit{Interest} and \textit{Data}. Each CCN node maintains three data structures; the \textit{Content Store (CS), Pending Interest Table (PIT) and Forwarding Information Base (FIB)}. CCN communication is consumer driven, i.e., a consumer sends \textit{Interest} packet towards the content source based on the information stored in the FIB. When a node receives an interest, it checks its local cache for the matching content. Otherwise, the node forwards the \textit{Interest} packet to the interface(s) based on the FIB table until the \textit{Interest} packet reaches a content source that can satisfy the interest. Intermediate nodes store the interests in the PIT so that the data can be sent back to the proper requester. In addition, PIT is used to suppress the forwarding duplicate interests over the same interface and provides response aggregations. CCN interests that are not satisfied within a reasonable amount of time are retransmitted. As CCN senders are stateless~\cite{jacobson2009networking}, the consumer is responsible for re-expressing interests if not satisfied.

Intermittently connected network topology or network disruption means a significant challenge for ICN deployment. For instance, name resolution may fail due to network disruptions, especially when the elements of the distributed resolution services are affected by network partitioning. Delay-tolerant networking~\cite{cerf2007delay} architectures are proposed for such scenarios, which are characterised by long delay paths, frequent unpredictable disconnections, and network partitions. Such architectures provide flexible and resilient protocols that build an opportunistic network on top of existing underlying Layer 2 and Layer 3 protocols. This is achieved through asynchronous communication along with the use of underlying Convergence Layer Adapters (CLA) (TCP, UDP, Bluetooth, etc.).

DTN is based on store-carry-and-forward models that utilise persistent storage that is distributed in the network. Data are cached in the network and are available for opportunistic transmissions. In particular, content based routing has been explored in DTN architectures~\cite{costa2006adaptive}. The multitude of the network interfaces in modern mobile devices allows DTN mechanisms to work in parallel with conventional ones. For instance, mobile users who are connected to the internet via the cellular interface, can also use the WiFi-direct interface to exchange messages with their neighbours. DTN architectures assimilate properties of ICN architectures and vice versa.

\begin{figure}[t]
	\centering
	\includegraphics[width=\columnwidth]{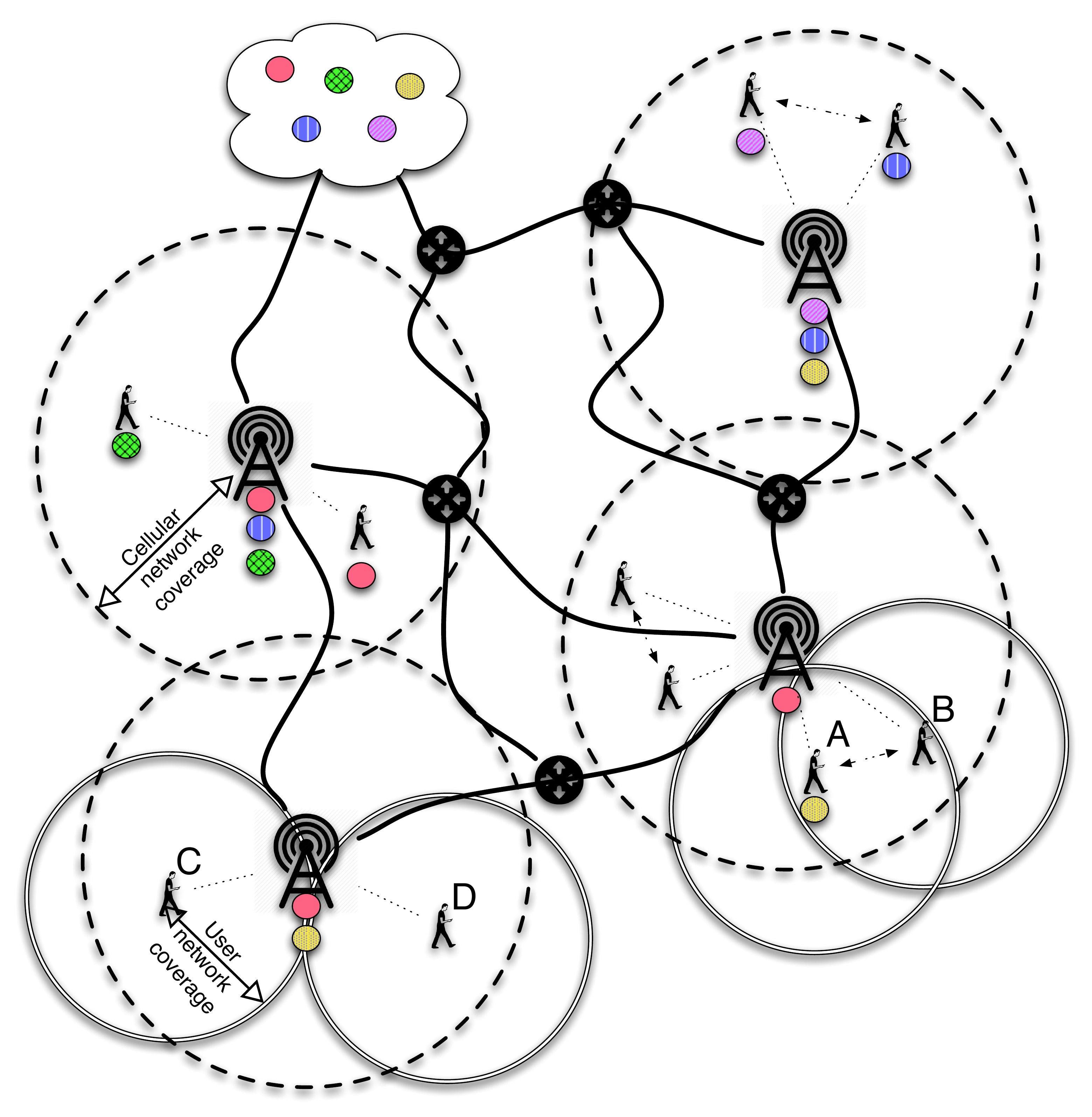}
	\caption{The examined ecosystem that combines Content Centric and Delay Tolerant architectures. \vspace{-0.5cm}}
	\label{fig:scenario}
\end{figure}

In this work, we adapt mechanisms from DTN networking to the ICN architecture in order to improve the efficiency of the content retrieval procedure of mobile users. In more detail, we consider the scenario where mobile users request \textit{content} via a CCN mechanism. These requests can be of many types, such as a single piece of data (e.g a request for the map of the current location of the user to Google Maps), a data stream (e.g., the homepage of a news website) or related to a specific type of information (e.g., opened restaurants close to the user). Some of the requests can be served more effectively by the cellular network but there are cases, like the third type of the request, that can be served locally. Such requests can potentially be served more effectively by nearby devices or by the cellular towers that have cached the requested content because another user requested that earlier. We considered the case, where contents are cached in the cellular towers but can also be requested from other nearby devices that have stored them. To achieve this, we modified the PIT table of the native CCN while operating on an opportunistic network. The modified PIT table stores the pending requester(s) information in the PIT table. The motivation behind this change is the fact that the original PIT table of the CCN keeps track of the arrival interfaces of the \textit{Interest} packets in a way that is not feasible in a highly dynamic network.

The contributions of this work can be summarised in the following list:

\begin{itemize}
  \item[\textit{(i)}] We explain what are the required modifications in the conventional CCN mechanism in order for it to be functional in a DTN environment. 
  \item[\textit{(ii)}] We propose a Content Centric DTN network architecture for mobile devices and introduce the required modification for the native CCN so that the native CCN can bridge with the Content Centric DTN protocol. The Content Centric DTN protocol operates independently of existing DTN routing protocols, i.e., DTN routing protocols run on top of the Content Centric DTN protocol. While designing our proposed architecture, we leverage the inherent properties of CCN and DTN architecture~\cite{cerf2007delay}.
  \item[\textit{(ii)}] We discuss the ways via which a mobile user can receive a requested content and show that the download time of a content can be decreased significantly via caching in the cellular access points and in other mobile devices.
  \item[\textit{(iv)}] We show that the underlying routing protocol does not have a substantial effect on the download time of contents due to the limited number of hops in the DTN. 
\end{itemize}

Figure \ref{fig:scenario} depicts the examined scenario where at any time mobile users are connected to the cellular network and are able to potentially communicate directly with other mobile users, depending on the distance between them and the underlying communication framework for the device-to-device communication. All the contents are stored in an origin server, which is located in a cloud infrastructure and can be cached to cellular access points and to mobile devices. Depending on the placement and the number of the cellular access points, the proportion of the content receptions from other mobile users differs significantly and, as we can see from our large scale simulations (Section \ref{sec:evaluation}), mobile users are able to successfully handle the requested contents in various different cases of routing schemes. 
After discussing the related work in the next section, we provide a more detailed explanation of the examined ecosystem in Section \ref{sec:model}. Next, in Section \ref{sec:alg}, we present the proposed protocol which is evaluated in Section \ref{sec:evaluation}. Finally, in \ref{sec:concl} we conclude the paper and list our future work.

\section{Related Work}

Based on the publish/subscribe paradigm, there exist numerous research efforts \cite{asadi2014survey} on device-to-device (D2D) communication in cellular networks, which is defined as direct communication between two mobile users without intervening Base Station (BS) or core network. This concept was first proposed in \cite{lin2000multihop}. Although D2D, from an architectural perspective, seems similar to Mobile Ad-hoc Networks, the key difference between these two is the involvement of the Cellular Access point. Casetti \textit{et al.} \cite{casetti2015content} presents content-centric routing in a D2D architecture based on Wifi Direct. The content-centric routing is based on two data structures: PIT of the native CCN and the Content Routing table (CRT). CRT provides the routing information to reach the content items. However, it is not feasible to maintain CRT and the PIT table in dynamic networks where mobile users provide intermittent connectivity. In contrast, our proposed scheme exploits the different PIT table which stores the requester(s) information instead of the arrival face of the original CCN so that the reverse path can be different from the forwarding path of the \textit{Interest} packets. Nevertheless, most recently, Garcia \textit{et al.} \cite{garcia2016ifip} have concluded that \textit{Interest} aggregation should not be an integral component of Content-Centric Networks and propose far smaller and more efficient forwarding data structures (e.g., CCN-DART \cite{garcia2016light}. 

Another similar effort has been proposed in \cite{helgason2010mobile} that allows wireless content dissemination between mobile nodes without relying on infrastructure support. The proposed architecture is based on the publish/subscribe paradigm. Their focus is mainly on implementation aspects based on 802.11 in ad-hoc mode. In contrast, our architecture is based on CCN and DTN architecture and hence there are many architectural differences between their effort and our proposal. Most recently, Liu et al. \cite{liu2017information} presents detailed descriptions on content routing based on ICMANET, and describes a concept model for content routing, and categorizes content routing into proactive, reactive and opportunistic types, then analyzes representative schemes, which can be referred to for the study of joint optimization between content routing and caching in ICMANET. There are also several research efforts in the DTN environment \cite{tysontowards}, \cite{trossen2016towards}. In \cite{tysontowards}, the author investigates the possibility of integrating the ICN and the DTN principles into a shared ICDTN architecture. Combining the ICN and the DTN has been demonstrated in a recent effort called RIFE architecture \cite{trossen2016towards}. The RIFE is a universal communication architecture that combines the publish/subscribe based POINT architecture \cite{point} and the DTN through a number of handlers for existing IP-based protocols (e.g., HTTP, CoAP, basic IP) which are mapped onto appropriate named objects within the ICN core. The IP endpoints are connected through the ICN using a gateway. In contrast, our proposed model exploits the DTN architecture in the native CCN architecture that results in a Multihop Cellular Network (MCN) \cite{lin2000multihop}. Amadeo et al. \cite{amadeo2016information} have discussed the potential of the ICN paradigm as a networking solution for connected vehicles. The authors have summarized ICN-VANETs relevant literature and presented the open challenges in this area. Nevertheless, the analysis of their work shows that the native design principles of ICN well match the main distinctive features of VANETs and the targeted wide set of future vehicular applications. The authors of \cite{saxena2017implementation} presents IP-based data DTN routing mechanisms using CCN on the sparsely-connected real vehicular testbed and validate the performance and usability of CCN over VANET. However, their proposed schemes have not considered the forwarding loop and duplicates at the content level while operating on IP-based routing mechanisms. Our proposed model operates independently of DTN routing and can detect the duplicates, and forwarding loop at content level.

User-centric data dissemination in DTNs has been widely explored from various points of view~\cite{sourlas2015information,costa2008socially, yoneki2007socio, lu2014information}. Authors of \cite{gao2011user} proposed a user-assisted in-network caching scheme, where users who request, download, and keep the content contribute to in-network caching by sharing their downloaded content with other users in the same network domain. Sourlas \textit{et al.} \cite{sourlas2015information} proposed an information-resilience scheme in the context of Content-Centric Networks (CCN) for the retrieval of content in disruptive, fragmented networks depending on the in-network caching of its attached user. The proposed scheme enhanced the Named Data Networking (NDN) router design as well as the \textit{Interest} forwarding mechanisms so that users can retrieve cached content when the content origin is not reachable. To achieve this, the authors introduce a new table, called Satisfied Interest Table (SIT), which keeps track of the Data packets that are forwarded to users. In case the content origin is not reachable, the proposed scheme exploits the cache of the other users following SIT entries. However, the proposed scheme performs well only if the users listed in the SIT entries are connected. In \cite{anastasiades2016information}, the authors present agent-based content retrieval on top of CCN which provides information-centric DTN support as an application module without modifications to CCN message processing. However, their proposed scheme may suffer from PIT bottleneck in delay tolerant environment. In contrast, our proposed scheme exploits the opportunistic communication of mobile users using DTN mechanisms. 

From a social-based point of view, the authors of SocialCast \cite{costa2008socially} proposed a routing framework that exploits the social ties among users for effective relay selection, while Yoneki et al. in \cite{yoneki2007socio} discussed the design of a publish-subscribe communication overlay based on the distributed detection of social groups by means of centrality measures. However, this routing mechanisms can be complementary to our proposed scheme, which operates independently of any routing algorithm. 
Lu \textit{et al.} at \cite{lu2014information} used the K-means clustering algorithm to build the social level forwarding scheme in order to reduce the transmitted messages. This approach raises several inevitable limitations: \textit{(i)} the interest may fail to reach the encountered node with the same social level that might have the content to satisfy the interest, \textit{(ii)} the request from the higher social level will never reach a content provider with a lower social level, \textit{(iii)} the proposed scheme cannot detect the routing loop of the \textit{Interest} packet and \textit{(iv)} the authors have not considered how to optimise similar interests from multiple users. These limitations are addressed in our solution.

D2D communication highly depends on the participation of mobile users in sharing contents. Mobile users may be selfish and would not be willing to forward data to others due to limited resources (e.g., memory, battery power). To handle this issue, a number of incentive mechanisms \cite{noura2016survey, jaimes2015survey, zhang2016incentives} has been proposed to motivate users to work in a cooperative way. D2D is still immature and faces many technical challenges and issues regarding aspects such as device discovery, relay selection, security and interference mitigation. The authors of \cite{jethawa2017incentive} presents an incentive mechanism for data centric message delivery in DTN that exploits the social relationships. This mechanism prevents users from becoming selfish and motivates them to relay the most popular content. Nevertheless, the incentive mechanisms are complementary to our proposed model and can be applied on top of our solution. In this work, we assume that all mobile users are participating in a cooperative way.

%Game \cite{xu2016game}

%zhang2015contract, wen2015quality,

\section{System Model}\label{sec:model}

\subsection{Preliminaries}

We analyse a CCN architecture where mobile users make a request for named data contents $c \in \mathcal{C}$. We consider a set of mobile users $\mathcal{M}$ that browse in a large scale metropolitan area and produce their requests for content\footnote{The terms user, node and device are used interchangeably depending on the context.}. A set of Cellular Access Points (CAPs) $\mathcal{A}$ are deployed in the area and we assume that at any time $t$ any mobile device $m \in \mathcal{M}$ is associated with one cellular access point and we denote this by $m_{a}(t) \in \mathcal{A}$, while any CAP $a$ has $\mathcal{N}_{a}(t)$ mobile devices associated with it at time $t$. 

Each CAP $a$ also operates as a CCN node by being connected to the fixed network and it maintains a \textit{Pending Interest Table} $P_a$ and a \textit{Forwarding Information Table} $F_a$. Also, part of its storage $S_a$ is used for caching contents and works as a \textit{Content Store}. The cache of each CAP is measured based on the proportion of total contents that it can store $S_a = \alpha_{\mathcal{A}}|\mathcal{C}|$, $0< \alpha_{\mathcal{A}} << 1$. In addition to the three traditional tables that are used in CCN architectures we add one table, motivated by the work of \cite{sourlas2015information}, which stores the satisfied interests. We denote that table with $D_a$. The entries of $D_a$ are of the form: $<${\tt content, user, time}$>$ and work in a similar way to the forwarding interest table, but with the difference that they keep who has satisfied its interest. In addition, we add another table that stores the \textit{Pending Requester Information table (PRIT)} that stores the requester information instead of the arrival interface. PRIT is used when the CAP receives requests from the DTN interface.

Any mobile node $m$ is able to communicate directly with its neighbours $\mathcal{N}_{m}(t)$, whose number depends on the mobility of the users, and the interface used for the connectivity between them\footnote{Bluetooth has a coverage radius of some tens of meters, WiFi-direct of a few hundreds and the soon-to-be-available LTE-direct is expected to have a coverage radius of half a kilometer.}. We also denote by $\mathcal{N}_{m}^{k}(t)$ the mobile users $m$ being able to communicate in $k>1$ hops, at time $t$. Similarly to the CAPs, each mobile node $m$ keeps three tables a \textit{Pending Requester Information Table} $P_m$, a \textit{Forwarding Information Table} $F_m$ and a \textit{Satisfied Interest Table} $D_m$ and has a \textit{Content Store}, $S_m$. The cache of each node is measured based on the proportion of total contents that it can store $S_m = \alpha_{\mathcal{M}}|\mathcal{C}|$. We assume that the storage capabilities of cellular access points is much higher than that of the mobile devices (e.g., some Terabytes compared to a few Megabytes), $0< \alpha_{\mathcal{M}} << \alpha_{\mathcal{A}} << 1$. Table~\ref{tab:notation} contains the introduced notation\footnote{To avoid listing the same variables for both mobile devices and CAPs we use $x$, $\mathcal{X}$ and $y$ (i.e. $x = \{m,a\}$, $\mathcal{X} = \{\mathcal{M},\mathcal{A}\}$ and $y=\{1,2,3\}$).}.

\subsection{Problem Formulation}

Each mobile user $m \in \mathcal{M}$ requests contents $c \in \mathcal{C}$ at the rate $r_{m}^{c}$. We use the vector $\textbf{r}^{c} \in \mathcal{R}_{+}^{|\mathcal{M}|}$ to denote all the request rates of all the mobile users for content $c$ and the zero norm\footnote{The zero norm of a vector equals to the non-zero elements of the vector.} of $\textbf{r}^{c}$, $||\textbf{r}^{c}||_{0}$ to indicate the number of the mobile users that are requesting content $c$. The request rate may depend on multiple factors, but in this work we consider only the popularity of the content $\pi_c$ and the profile of the user $u_m$, which indicates the probability of a mobile user requesting each content. We denote the profiles of all users with the vector $\textbf{u}$. So the request rate of content $c$ by user $m$ is given by: 
\begin{equation}
	r_{m}^{c} = u_m \cdot \pi_c
\end{equation}
The service rate of an expressed interest from a user $m$ and a content $c$ depends on the popularity of the content $\pi_c$ and the content placement strategy that will be explained in detail in Section \ref{sec:alg}. An interest in a content from a mobile user can be served in three ways: 

\subsubsection{Core Network}
The mobile device, via the cellular network, sends the \textit{Interest} packet and the content is retrieved in the traditional CCN way from the Content Store of any intermediate node or from the server of origin, where the content was initially placed upon its creation. At any time $t$ there exists at least one node that has the required content. In such a case, the service rate of content $c$ is denoted by $s_{N_{1}}^{c}$ and depends on the popularity of the content and the characteristics of the network (load, bandwidth, etc) and the caching policy (e.g., LRU, FIFO, LFU), $c_{N_{1}}$. Without loss of generality we assume: 
\begin{equation}
	s_{N_{1}}^{c}(t) = c_{N_{1}}(t) \cdot \pi_c
\end{equation}

\subsubsection{Cellular Access Point}
The mobile device downloads the cached content from the cellular tower because another user had requested the content earlier. Given that the available cache of each cellular tower is limited compared to the storage size of the server of origin, the cached contents are limited ($\alpha_{\mathcal{A}}|\mathcal{C}|$) but, depending on the caching policy, can achieve a high hit rate due to the popularity distribution of the contents and the spatial skewness \cite{Fayazbakhsh:2013:LPM:2486001.2486023}. In that case, we denote the service rate with: 
\begin{equation}
	s_{N_{2}}^{c}(t) = \alpha_{\mathcal{A}} \cdot c_{N_{2}}(t) \cdot \pi_c
\end{equation}

\subsubsection{Delay Tolerant Network}
The mobile device gets the content from another mobile device via a single-hop or a multi-hop path. The number of the hops depends on \textit{(i)} the physical distance between the users, \textit{(ii)} the number of the users and \textit{(iii)} the popularity of the content. Popular contents are more probably found closer to the user who initiated the request. Although mobile devices are not able to cache many content items, the social relationship between mobile users that have, with high probability, similar mobility patterns, makes it probable for two socially close mobile users to express interest in similar items \cite{hui2011bubble}. In that case, we denote the service rate with: 
\begin{equation}
	s_{N_{3}}^{c}(t) = \alpha_{\mathcal{M}} \cdot c_{N_{3}}(t) \cdot \pi_c \cdot  \prod_{m \in \mathcal{N}_{m}(t)}u_{m} 
\end{equation}

\begin{table}[t]
%\normalsize
	\centering
	\caption{Notation Table}
	\label{tab:notation}
	\begin{tabular}{m{0.1\columnwidth}  m{0.75\columnwidth}}
	%	Symbol & Meaning \\ \toprule
	\toprule
		$\mathcal{C}$			& set of available contents \\ \midrule
		$\mathcal{M}$ 			& set of mobile users   \\ \midrule
		$\mathcal{A}$ 			& set of cellular access points \\ \midrule
		$\alpha_{\mathcal{X}}$ 	& cache capacity of mobile users $\mathcal{M}$ or CAPs $\mathcal{A}$ \\ \midrule 
		$\mathcal{N}_{x}(t)$ 		& mobile users accessible by user $m$ or CAPs $a$ at time $t$ \\ \midrule 
		$P_x$				& Pending Interest Table of user $m$ or CAP $a$ \\ \midrule
		$F_x$				& Forwarding Infromation Table of user $m$ or CAP $a$ \\ \midrule
		$S_x$ 				& Content storage of user $m$ or CAP $a$ \\ \midrule
		$D_x$				& Satisfied Interest Table of user $m$ or CAP $a$ \\ \midrule
		$\pi_c$				& Popularity of content $c$ \\ \midrule
		$u_m$				& Content request profile of mobile user $m$ \\ \midrule
		$r_{m}^{c}$			& Request rate of content $c$ from mobile user $m$\\ \midrule
		$s_{N_{y}}^{c}(t)$		& Service rate of content $c$ at time $t$ through network $N_{y}$ \\ \bottomrule 
	\end{tabular}
	\vspace{.2cm}
\end{table}

We employ a Markov process $\{ X_{c}(t), 0 \leq t < \infty \}$ with stationary transition probabilities that shows the number of the nodes in the whole ecosystem (mobile users, cellular access points, the server of origin as well as the network components such as switches that are part of the CCN ecosystem that have the content $c$ in their caches). If at any time $\tilde{t}$, $X_{c}(\tilde{t}) = 0$, this will mean that the content is not available at all, which can be true only in the case of a very unpopular item that is not cached in any node and the server of origin is not accessible because of network partitioning. However, although this is not realistic, we can use the Markov process as a birth-death process with a single absorbing state, which we define to be $X_{c}(t) = 0$ in order to then use the absorption time formula~\cite{taylor2014introduction} that includes the cost parameters for each type of network as an objective function to optimise. In more detail~\cite{taylor2014introduction}:
\begin{eqnarray}
T_{n}^{c} &=& \sum_{i=1}^{\infty} \frac{1}{\lambda_{i}^{c} \rho_{i}^{c}} + \sum_{k=1}^{n-1}\rho_{k}^{c} \sum_{j=k+1}^{\infty}\frac{1}{\lambda_{j}^{c} \rho_{j}^{c}}, \\ 
&\text{if}& \hspace{2 cm} \sum_{i=1}^{\infty}\frac{1}{\lambda_{i}^{c} \rho_{i}^{c}} <\infty
\end{eqnarray}
and $T_{n}^{c} = \infty$, if $\sum_{i=1}^{\infty}\frac{1}{\lambda_{i}^{c} \rho_{i}^{c}} = \infty$, where: $\lambda_{n}^{c}$ is the birth rate of the process at state $n$, $\mu_{n}^{c}$ is the death rate and 
\begin{equation}
	\rho_{n}^{c} = \prod_{i=1}^{n} \frac{\mu_{i}^{c}}{\lambda_{i}^{c}}
\end{equation}

The birth rate of the process at state $n$ and for content $c$, $\lambda_{n}^{c}$, depends on the request rates for the examined item of each of the users $r_{m}^{c}$. 
\begin{equation}
	\lambda_{n}^{c} \sim \sum_{ \substack{ m \in \mathcal{M} \\ || \textbf{r}^{c}||_{0} = n}} r_{m}^{c}
\end{equation}
while the death rate depends on the type of the service rate and the caching policies. The required time for the Markov process to reach the absorption state depends on the initial state and the difference between the service rates and the request rates. 

The service rate depends on the probability of a content being placed close to the mobile users that generate requests for it. The probability of a content $c$ being cached in the CAP which mobile user $m$ is associated with at time $t$, $m_{a}(t)$ is: 
\begin{equation}
p_{m_{a}}^{c} (t)\coloneqq P[c \in S_{m_{a}(t)}]
\end{equation}
and the probability of $c$ being stored in at least one of $m$'s neighbours is  
\begin{equation}
	p_{\mathcal{N}_{m}}^{c}(t) \coloneqq 1 - \sum_{j \in \mathcal{N}_{m}(t)} P[c \notin S_{j}(t)],
\end{equation}
while for $K$ hops away from $m$, the probability of $c$ being cached is: 
\begin{equation}
	p_{\mathcal{N}_{m}^{K}}^{c}(t) \coloneqq 1 - \sum_{j \in \mathcal{N}_{m}^{K}(t)} P[c \notin S_{j}(t)]. 
\end{equation}
So the probability for a mobile user not being able to retrieve $c$ from the access point that he or she is associated with and from any mobile user whose distance is at most $K$-hops is \footnote{Small values of $K$ are enough for successful content discovery~\cite{Wang:2015:PUS:2810156.2810162}.}: 
\begin{equation}
	p_{m}^{c}(K,t) \coloneqq 1 - p_{m_{a}}^{c} (t) - p_{\mathcal{N}_{m}}^{c}(t) - \sum_{k=2}^{K} p_{\mathcal{N}_{m}^{k}}^{c}(t).
\end{equation}
 We also define the probability of a content $c$ being cached in the cellular access point $a$ of at least one of the mobile devices that are associated with $a$: 
\begin{equation}\label{eq:prob_c_in_a}
	p_{a}^{c}[\mathcal{N}_{a}(t),t] = 1 - P[c \notin S_{a}(t)]\cdot \prod_{j \in \mathcal{N}_{a}(t)} P[c \notin S_{j}(t)]
\end{equation}

The size of the Content storage in the CAPs and mobile devices, and more specifically the proportion of the total items they can store is what affects $p_{m}^{c}(K,t)$ and $p_{a}^{c}[\mathcal{N}_{a}(t),t]$. Another determinant parameter is the number of mobile users that are associated with the same access point as the user that requested a content item and, consequently, the diversity in the subset of the objects that are cached in all these devices. We denote with $\mathcal{C}_{a}(t) \subset \mathcal{C}$ the set of the content items that are cached in at least one device that is accessible from CAP $a$ or are cached in $a$. Then equation (\ref{eq:prob_c_in_a}) can be expressed shortly as $P[c \in \mathcal{C}_{a}(t)]$. 

 Next, in Section \ref{sec:alg}, we present a protocol that determines which contents should be cached in each device and for how long. The protocol is designed to consider highly dynamic mobile users with limited resources as well as the static access points that operate as the glue between the dynamic users and the fixed infrastructure.

\begin{figure}[t]
	\centering
	\includegraphics[width=\columnwidth]{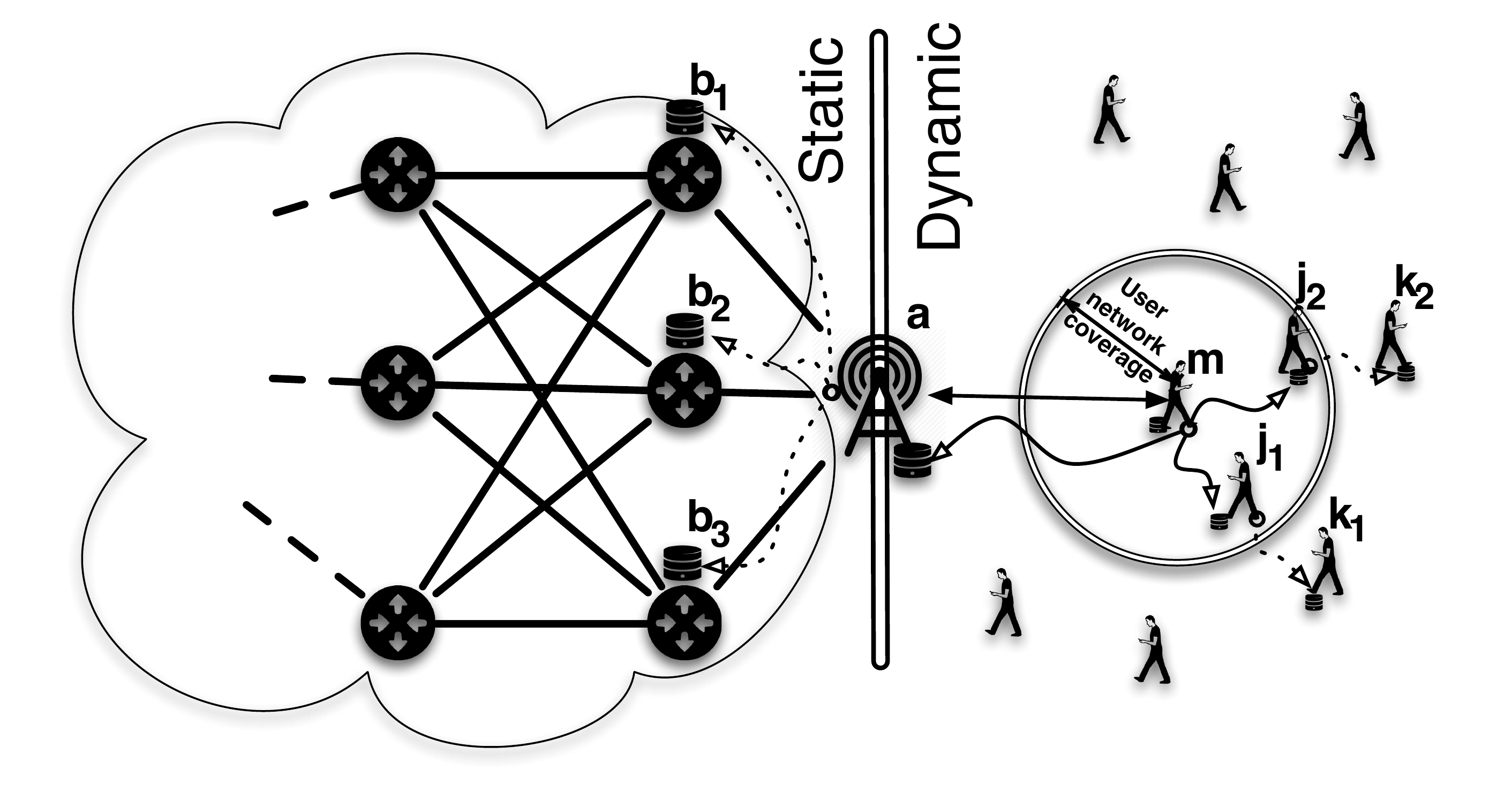}
	\caption{The cellular access points connect the CCN with the static links between the nodes to the highly dynamic and unpredictable DTN.}
	\label{sec:scenario}
\end{figure}

\section{Protocol}\label{sec:alg} 

The original design of CCN is based on the fact that the multiple network interfaces can be integrated via the mechanism of the forwarding information base~\cite{jacobson2009networking}. Each entry on the FIB points to a list of interfaces that can be used to forward \textit{Interest} packets towards the desired content producer. At this point, the traditional CCN can be combined with DTN network protocols, as presented in figure \ref{sec:scenario}. The integration of DTN architecture with the native CCN architecture results in a Multihop Cellular Network (MCN) \cite{lin2000multihop}. The general concept of MCN comprises a cellular network in which user devices can communicate with each other, either via means of a conventional cellular mode or via means of direct
D2D communication if they are mutually reachable. To enable this paradigm, the functionalities of the proposed protocol can be decomposed into three parts:

\textbf{(1) The control plane} that performs packet (\textit{Interest/Data}) management. The control plane is implemented on top of the DTN mechanisms, and its functionalities are responsible for performing specific actions based on the packet type (\textit{Interest/Data}). To achieve this, the control plane inserts the meta-information in the DTN messages.

\textbf{(2) The forwarding plane} that consists of two parts and, depending on the type of the node, it can be either the native \textit{CCN forwarding} or the\textit{ DTN forwarding (Store-carry-and-forward)}. This module provides an interface between the CAP and the mobile nodes so that the CAP can hand over the packet to a mobile node. The mobile node exploits DTN architecture to forward the packets in D2D fashion while operating in an opportunistic network without the intervention of the cellular network. The CAP includes a separate PIT table called the Pending Requester Information Table (PRIT) which stores the requester(s) information instead of the arrival faces of the \textit{Interest} packets. The mobile nodes only use our proposed architecture while operating in a DTN environment, i.e., the control plane is implemented on top of the DTN forwarding plane and enables the host centric DTN to perform in content centric fashion. To bridge between CCN and DTN, each message carries meta-information of the CCN mechanism that assists the content centric operation in DTN environment.  

\textbf{(3) The routing decision engine} is the process by which one router sends packets to another router by means of routing protocols which decide the appropriate path for the packet. The routing protocol assists the router in choosing the best path out of many paths. The routing decision engine operates independently on top of our proposed model.

The proposed protocol deals with two control decisions: 

\begin{enumerate}
\item \textit{Request/Response Processing}: Although the CAPs operate as conventional CCN nodes regarding the forwarding and the routing of \textit{Interest} and \textit{Response} packets, it is not the same for mobile users. Whenever a mobile user of a CAP receives a content request or a content response, there is the question of what actions should be taken?
\item \textit{Content Management}: Given that a mobile user or a CAP has a content item, should it store it in the content store or drop it? The CAPs have higher storage capabilities than the mobile users, but still they can not cache all the available contents. 
\end{enumerate}

\subsection{Request Processing}

In the relatively static CCNs, the \textit{Interest} packets are propagated as upstream towards the potential data sources, while leaving a trail of \textit{bread crumbs} for the matching data packets to follow back to the original requester(s). On the other hand, in dynamic environments the nodes are mobile and the connections are intermittent, which means that it is not feasible to keep track of the changes in the network topology. Unlike the conventional PITs in CCN, mobile users keep the address information of the requester(s) in the $Pending\ Requester\ Information\ Table\ (PRIT)$ so that they can forward similar content towards potential requester(s). PRIT is also used to detect forwarding loop and aggregate the similar interests. Mobile users exploit the $Satisfied\ Request\ Information\ Table\ (SRIT)$ to remember all the satisfied interests of the requester(s) so that it can provide information on the potential content source for the similar interests in future. %and use the SRIT to remember all intermediate nodes that forwarded the interest packet. 
By doing that, an intermediate node having an entry matching with the interest packet in the SRIT can forward the \textit{Interest} packet towards those potential content provider(s). The CAP acts similarly to a mobile node if it receives the \textit{Interest} packet from the DTN interface. Nevertheless, if the CAP has FIB entry for this \textit{Interest}, it can also apply the native CCN mechanism. The overview of the request processing is presented in Algorithm~\ref{interest}.

\begin{algorithm}[t]
\algsetup{linenosize=\small}
\small
\caption{Processing Interest Message}
\label{interest}
\begin{algorithmic}[1]

\STATE $key \leftarrow [Interest]$
\IF{$key\ in\ Local\ Cache $}
   \STATE $content \leftarrow Cache(key)$ 
 \ENDIF 
 
\IF {$content \neq NULL$}
 \STATE $response \leftarrow createResponse(content)$
  \IF{$ current\_node $ = $mobile\ user$}
    \STATE $requester \leftarrow [Interest] $
    \STATE $Send\ response\ to\ requester\ following\ PRIT$
   \ELSE
     \STATE $Send\ response\ following\ PIT\ breadcrumb$
   \ENDIF   
\ELSE 
   \IF{$ current\_node $ = $mobile\ user$}
     \STATE $satisfied\_req\_provider \leftarrow lookup\_SRIT(Interest)$
     \IF {$satisfied\_req\_provider \neq NULL $}
       \STATE $Send\ Interest\ to\ satisfied\_req\_provider\  $
      \ELSE
       \STATE $pending\_requester \leftarrow lookup\_PRIT(Interest)$
       \IF{$requester \in pending\_requester$}
         \STATE $ drop\ the\ interest\ packet$
        \ELSE 
          \STATE $Add\ requester\ to\ PRIT\ table$ 
          \STATE $forward\ the\ Interest\ to\ next\ Hop $
        \ENDIF  
     \ENDIF
   \ENDIF 
  \IF{$ current\_node $ = $CAP$}
     \STATE $ FIB\_entry \leftarrow native\_CCN\_mechanism(Interest)$
     \STATE $satisfied\_req\_provider \leftarrow lookup\_SRIT(Interest)$
     \IF{$FIB\_entry$ = $NULL $}
        \IF {$satisfied\_req\_provider \neq NULL $}
          \STATE $Send\ Interest\ to\ satisfied\_req\_provider\ $
        \ELSE
         \STATE $pending\_requester \leftarrow lookup\_PRIT(Interest)$
          \IF{$requester \in pending\_requester$}
           \STATE $ drop\ the\ interest\ packet$
          \ELSE 
          \STATE $Add\ requester\ to\ PRIT\ table$ 
           \STATE $forward\ the\ Interest\ to\ mobile\ user $
           \ENDIF
        \ENDIF
       \ENDIF 
   \ENDIF  

%\STATE $Add\ source\ EID\ to\ PRIT\ table$
%\STATE $forward\ the\ Interest\ to\ next\ Hop $

\ENDIF

\end{algorithmic}
\end{algorithm}

%we are able to handle the case where a mobile node meets the requester of the content and delivers it to her without the need of forwarding it to the intermediate nodes. However, we also need to consider all the ids of the potential requesters for the same content. 

On the reception of an \textit{Interest} packet, a mobile node initially searches in its Content Store and if there is no match, the node checks its SRIT table to verify if there is any entry matching with the \textit{Interest} packet. If any matching is found, the node forwards the request towards those potential content provider(s) from SRIT. The node also enters the \textit{Interest} packet in the PRIT table. The PRIT is used to keep track of the IDs \footnote{Without loss of generality we assume that the id of user $m$ is $m$.} of the interest(s) creators that are used as destinations in the response packets. In more detail, upon the reception of an \textit{Interest} packet the mobile node checks its PRIT. If there is an older entry for the same content, it updates the entry only if the requester is different, otherwise it drops the \textit{Interest} packet. On the other hand, the CAP first applies the native CCN mechanism, i.e., it searches the content store to verify if it can satisfy the request. If there is a match, the CAP sends the content back to the requester. If no matching is found, the CAP forwards the request further, based on the information of the FIB. The CAP can also forward the request to the mobile node which runs our proposed architecture. Before forwarding the request to the mobile node, the CAP will store the requester information in the PRIT table, but only if it receives the request from the DTN interface. 

\begin{figure}[t]
	\centering
	\includegraphics[width=\columnwidth]{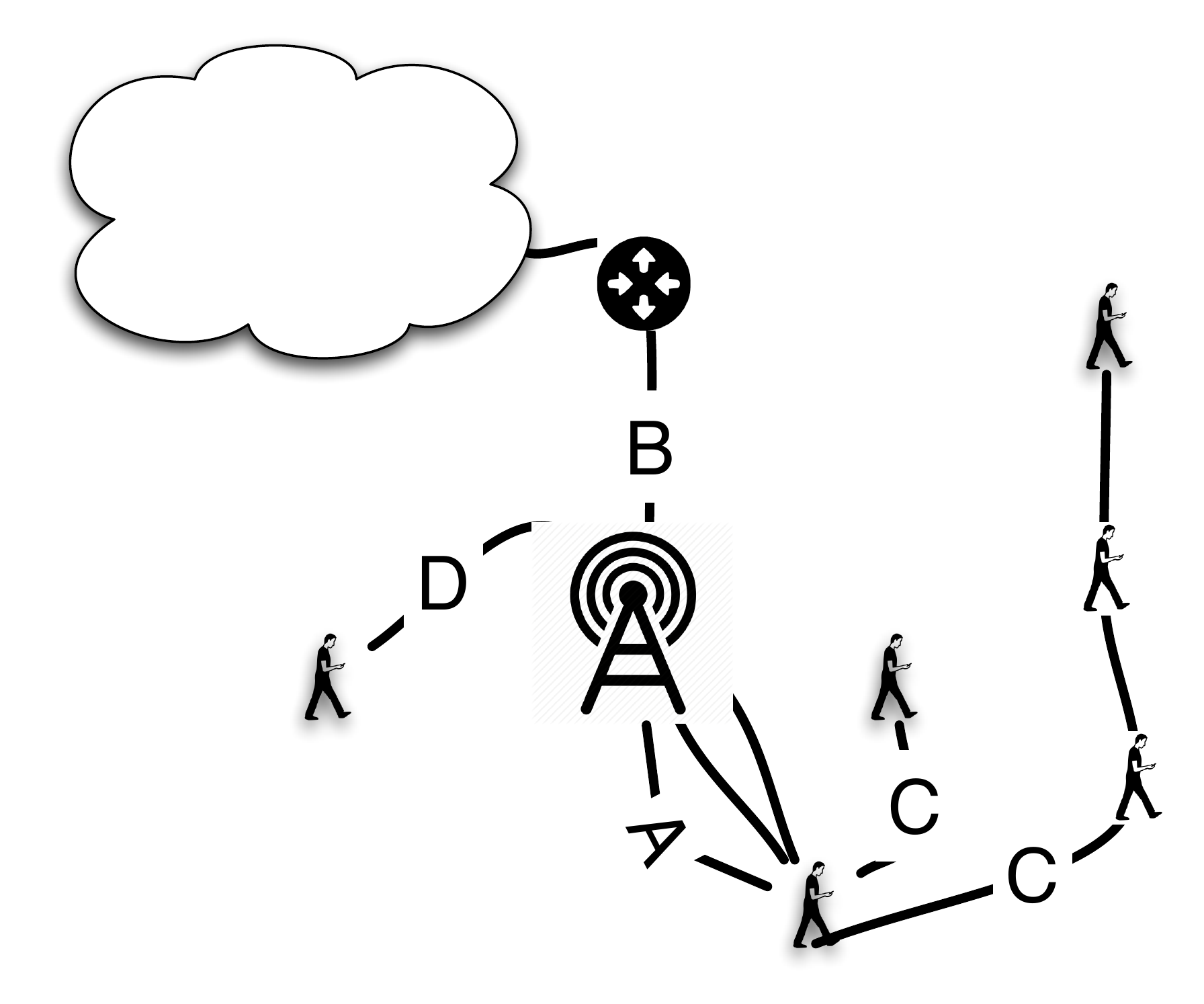}
	\caption{The four potential ways via which a mobile user can get the requested content item.  \vspace{-0.3cm}}
	\label{fig:downloadpaths}
\end{figure}

Regardless of the total number of users, our proposal does not spread the \textit{Interest} packets all over the ecosystem because it is inefficient and not worthwhile doing since the mobile nodes are submitting their requests in parallel to both the CAPs they are connected to and their neighbouring mobile devices. More importantly, the respective CAPs inform the mobile nodes whether there exists another mobile node that has the requested content in the same cell, and depending on the level of the assistance from the CAPs, as will be discussed in the next section, the mobile nodes can either receive their request via a multi-hop-but-short path from another node in the same cell, or via a two hop path with the help of the CAP. So, a request as shown in Figure \ref{fig:downloadpaths} can be served in four ways: \textit{(A)} from the Content Store of the associated CAP, \textit{(B)} via the associated CAP that retrieved the content from the conventional CCN network, \textit{(C)} from another mobile node that sent the content via a multi-path among the other mobile nodes, and \textit{(D)} from another mobile node that sent the content to the CAP, which then forwarded the content to the requester.

\begin{algorithm}[t]
\algsetup{linenosize=\small}
\small
\caption{Processing Response Packet}
\label{responsealg}
\begin{algorithmic}[1]
\IF {$ current\_node\ is\ mobile\ user$}
  \STATE $ destination\_id \leftarrow [Response]$
  \STATE $content\_provider \leftarrow [Response] $
  \STATE $insert\ content\_provider\ in\ SRIT\ table $
 \IF {$current\_node\ is\ the\ destination $}
   \STATE $notify\ application$
   \STATE $key \leftarrow [Response]$
   \STATE $pending\_requester \leftarrow lookup\_PRIT(key) $
   \IF {$ pending\_requester\ is\ empty $}
     \STATE $drop\ the\ packet$
     \RETURN
    \ELSE 
      \STATE $ forward\ response\ to\ pending\_requester $
      \RETURN 
    \ENDIF  
 \ENDIF  
\ENDIF

\IF {$ current\_node\ is\ CAP$}
  \IF{$ response\ is\ received\ from\ DTN\ interface$}
    \STATE $key \leftarrow [Response]$
    \STATE $content\_provider \leftarrow [Response] $
    \STATE $insert\ content\_provider\ in\ SRIT\ table $
    \STATE $pending\_requester \leftarrow lookup\_PRIT(key) $ 
    \IF {$ pending\_requester\ is\ empty $}
      \STATE $drop\ the\ packet$
     \ELSE  
       \STATE $ forward\ response\ to\ pending\_requester $ 
     \ENDIF
   \ELSE
     \STATE $follow\ the\ native\ CCN\ mechanism$ 
    \ENDIF
 \ENDIF    

%\STATE $add\ content\ to\ opportunitistic\ cache$
%\STATE $Forward\ response\ to\ next\ hop$

%\RETURN $P$
\end{algorithmic}
\end{algorithm}

\subsection{Response Processing} 

Algorithm~\ref{responsealg} presents an overview of the response processing on a  network node. When the \textit{Interest} packet reaches a node having content matching with the \textit{Interest} packet, the node constructs a response packet with the content and sends it back to the originator of the request. If the intermediate node is a mobile node, the node checks the PRIT table and removes the entry if there is a match for the response packet. If the PRIT entry has the information on multiple requesters, the intermediate node adds all the source IDs of those requesters to the response packet as meta-information. If the intermediate node does not find any matches in the PRIT table, it simply forwards the response packet to the next best contact. Subsequently, if the response packet reaches the target node, it checks the meta-information to verify if there is any other pending requester(s) who requested this content. If there exists no pending requester information, the recipient node drops the packet to avoid further transmission by the DTN mechanism. Otherwise, if the node finds other pending requesters, it will forward the response to those pending requesters. If the meta information has multiple pending requesters, the node adds one requester as the destination address for the response and other requester(s) as meta-information. If the intermediate node is the CAP, it checks both PIT and PRIT to forward the response in an appropriate manner. If the CAP finds a match in its PIT, it follows the native CCN mechanism. A match in PRIT follows our proposed scheme.

\subsection{Content Management}

Mobile nodes can provide storage memory depending on their resource availabilities and policies. Using its storage memory, a mobile node can serve as the network medium to share the content. Furthermore, this cache can also be used by store-carry-and-forward based DTN protocols. However, the persistent storage of the DTN protocol keeps the message until the successful delivery of the message to the next best opportunistic contact. In our proposed architecture, the storage memory of mobile node keeps the response packet to satisfy the future requests. However, only in the ideal case is the mobile node's storage big enough to store all received content. Under the assumption that a mobile node can only store a small proportion of the total contents, a caching policy is required.

Additional information, such as the popularity distribution of the contents and the request profile of the mobile users, can be used by a caching policy to determine the probability of each content being requested by the user and, based on that, a decision is made whether a newly received content be dropped or replaced with the unpopular one, given that there is no available space in the mobile node. However, the CAP has information about the stored content in the whole cell that is not utilised. We utilise this additional information by defining the expected retrieval cost of each content by combining this information: \textit{(i)} the popularity of each content item, \textit{(i)} the profile of the users and the \textit{(i)} estimated time required to retrieve the content via one of the aforementioned ways. This costs are calculated with the assistance of the CAPs, which is able to recommend a mobile node on whether to keep an item or not.

\section{Evaluation}\label{sec:evaluation}

We evaluate our proposed architecture using the Opportunistic Network Simulator (ONE) \cite{keranen2009one}. The goal of our evaluation is to investigate the performance of our proposal in terms of \textit{(i)} Average end-to-end delay,  \textit{(ii)} service load ratio, \textit{(iii)} packet drop ratio and \textit{(iv)} Traffic split. Table \ref{tab:metrics} contains a description of each of these metrics. 

\begin{table}[t]
%\normalsize
	\centering
	\caption{Performance Metrics}
	\label{tab:metrics}
	\begin{tabular}{m{0.2\columnwidth}  m{0.65\columnwidth}}
	\toprule
Average end-to-end delay	& The average time passed to receive a content in response to a request. \\ \midrule
Packet drop					& The amount of packets (Interest/Data) that is effectively suppressed by the content router.\\ \midrule
Traffic split			& The service rate, i.e., the amount of requests processed by different types of nodes in the network: mobile user, CAP, content source.  \\ \midrule
Service load 		& The amount of requests processed by the content provider. 
\\ \bottomrule
	\end{tabular}
	
\end{table}

The ONE simulator contains map data of the Helsinki downtown area (e.g., roads, tram routes and pedestrian walkways) and various Map-based Movement models: \textit{(1)} Random Map-Based Movement, \textit{(2)} Shortest Path Map-Based Movement, and \textit{(3)} Routed Map-Based Movement. We employ the Shortest Path Map-Based Movement since it is more realistic because the mobile users, after choosing a destination point on the map, follow the shortest path to that point from their current location. The destination point is chosen randomly from a list of Points of Interest (POI), which includes popular real world destinations (e.g., shops, restaurants, tourist attractions). The simulation area approximately is 20km$^2$.

In the simulation, we considered mobile users that are either walking at a speed that is in the range of 1.8 kilometres per hour to 5.4 kilometres per hour or driving a car or using the tram. We categorised the mobile users into two groups: \textit{(i)} requesters and \textit{(ii)} intermediate users. The requesters were 10 and the intermediate users were 150. All of them were divided into four different groups and assigned with different probabilities of choosing the next group specific POI or random places to visit. Regarding the content generation, we considered content generated by 10 other mobile users or from non-mobile content generators (e.g., a news website). Apart from the mobile users, we also considered 30 CAPs that have caching capabilities.

None of the users had any content in the beginning of the simulation, but whenever one requester imposed a request on a CAP, the content was retrieved from the content provider in the cloud if it had not already been cached from a previous request, and delivered to the requester. The simulation time was 5 days and we used the first day as a warm up phase. All the details of the simulation parameters are listed in Table \ref{sec:SimulationParameters}.

\begin{table}[t]
	\centering
	\caption{Simulation Parameters}
	\begin{tabular}{l l}   
		Parameter & Value \\ \toprule 
		Simulation Duration & 5 days (432000s) \\ 
		Number of Requesters & 10 \\
		Time interval of generating Interests & 5min \\ 
		Number of Relay Nodes & 160 \\
		Number of Access Points & 30 \\
		Cache of mobile users & 10 items \\
		Cache of Access points	& 50 items \\
		TTL value & 500s\\
		Transmission range of Access Points & 100m \\ 
		Transmission speed of Access Points & 10Mbps. \\ 
		Transmission range of Mobile devices & 10m \\ 
		Transmission speed of Mobile devices  & 2.5 Mbps.  \\ \bottomrule 
\end{tabular}
\label{sec:SimulationParameters} \vspace{-0.3cm}
\end{table}

\subsection{DTN Routing}\label{sec:dr}

The Content-Centric functionalities of our proposal are routing independent, and for that reason we examine the performance of our proposal in four different cases regarding the routing strategies: \textit{(i)} Epidemic~\cite{vahdat2000epidemic}, \textit{(ii)} Spray-and-Wait~\cite{spyropoulos2005spray}, \textit{(iii)} First contact~\cite{jain2004routing} and \textit{(iv)} a hybrid one that works like the Epidemic in the forwarding step until reaching the destination and also like the Spray-and-Wait in the reverse path creation step. Epidemic routing has no limitation on generating copies for each message. In this routing scheme, each node carries a list of all messages whose delivery is pending. Whenever a node encounters another node, they exchange all that messages that are not common in their list. 
Spray-and-Wait generates a limited number of copies for every message and spreads initially. If a node does not find the destination in the spray phase, it waits for the destination to perform direct transmission. In our experiment, Spray-and-Wait generated 10 copies for every message in the spray phase. First contact generates only one copy per message. The hybrid one sprayed \textit{Interest} packets (limited to 10 copies) until the request reached the \textit{content providers} and then used the Spray-and-Wait routing to deliver the content back to the requester.

\subsection{Query Distribution}\label{sec:dist}
We generate user interest based on the available contents $\mathcal{C}$, which we assume are 1000 (i.e. $|\mathcal{C}| = 100$). We assume that the $i$-th content $c_{i}$ is the $i$-th most popular one $\pi_i < \pi_j$ $\forall i \leq j$. The users' request profiles are randomly generated via the uniform distribution.  
Content popularity is correlated with user requests \cite{liu2005client} and follows the well-known Zipf distribution \cite{breslau1999web}. In this work we consider two cases for the content popularity: \textit{(i)} uniform and \textit{(ii)} Zipf on initialising $\pi_c \forall c \in \mathcal{C}$. For the Zipf distribution we initialised the parameter to 1 and the normalizing constant to 0.2. 

\begin{figure}[t]
\centering
\begin{subfigure}{\columnwidth}
	\includegraphics[width = \columnwidth]{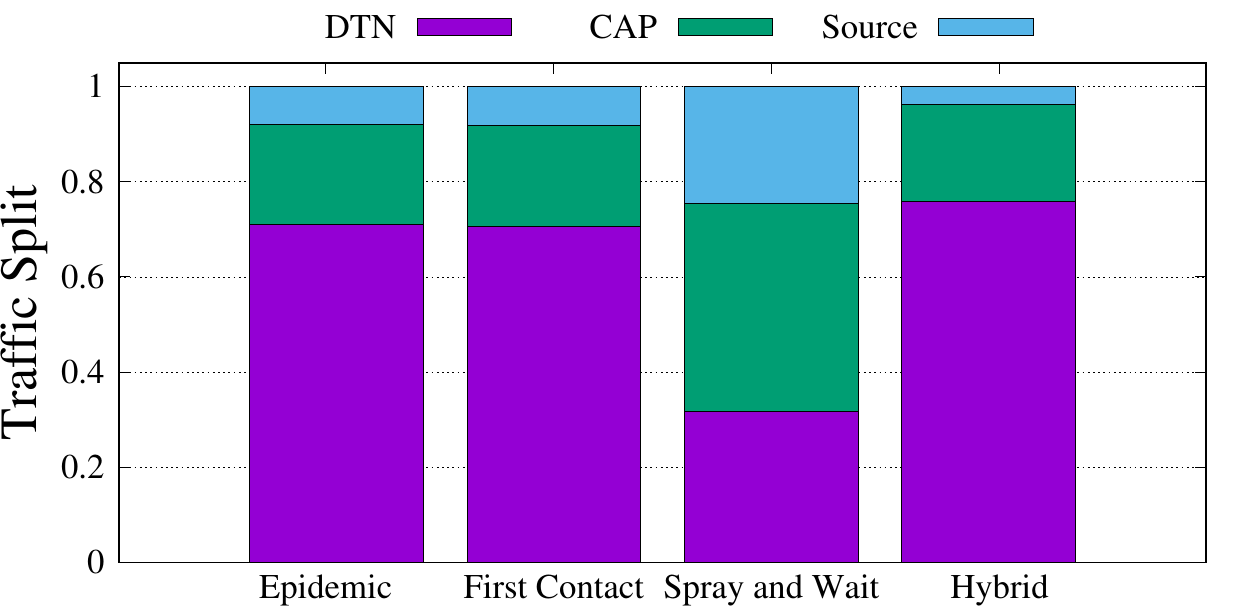}
	\caption{Uniform content popularity}
	\label{fig:trafficSplitUn}
\end{subfigure}
\begin{subfigure}{\columnwidth}
	\includegraphics[width = \columnwidth]{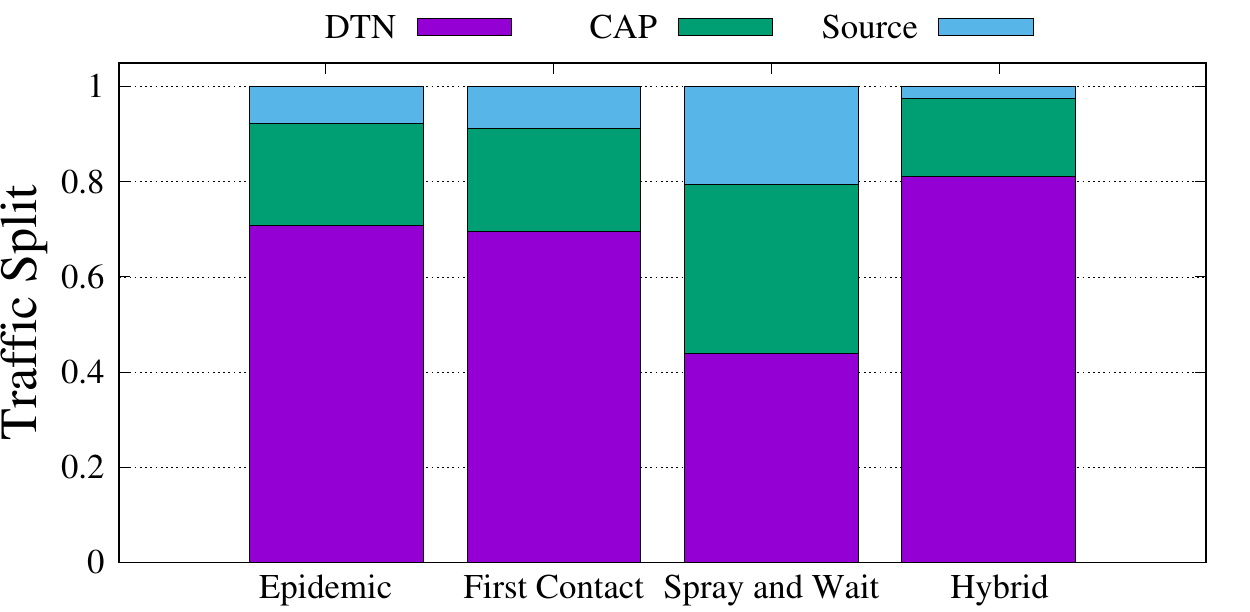}
	\caption{Zipf content popularity}
	\label{fig:trafficSplitZipf}
\end{subfigure}
\caption{The service rate, i.e., the distribution of the content responses from each type of source for each DTN routing protocol. \vspace{-0.3cm}}
\label{fig:trafficSplit}
\end{figure}

\begin{figure}[t]
\centering
\begin{subfigure}{\columnwidth}
	\includegraphics[height = \columnwidth,angle=270]{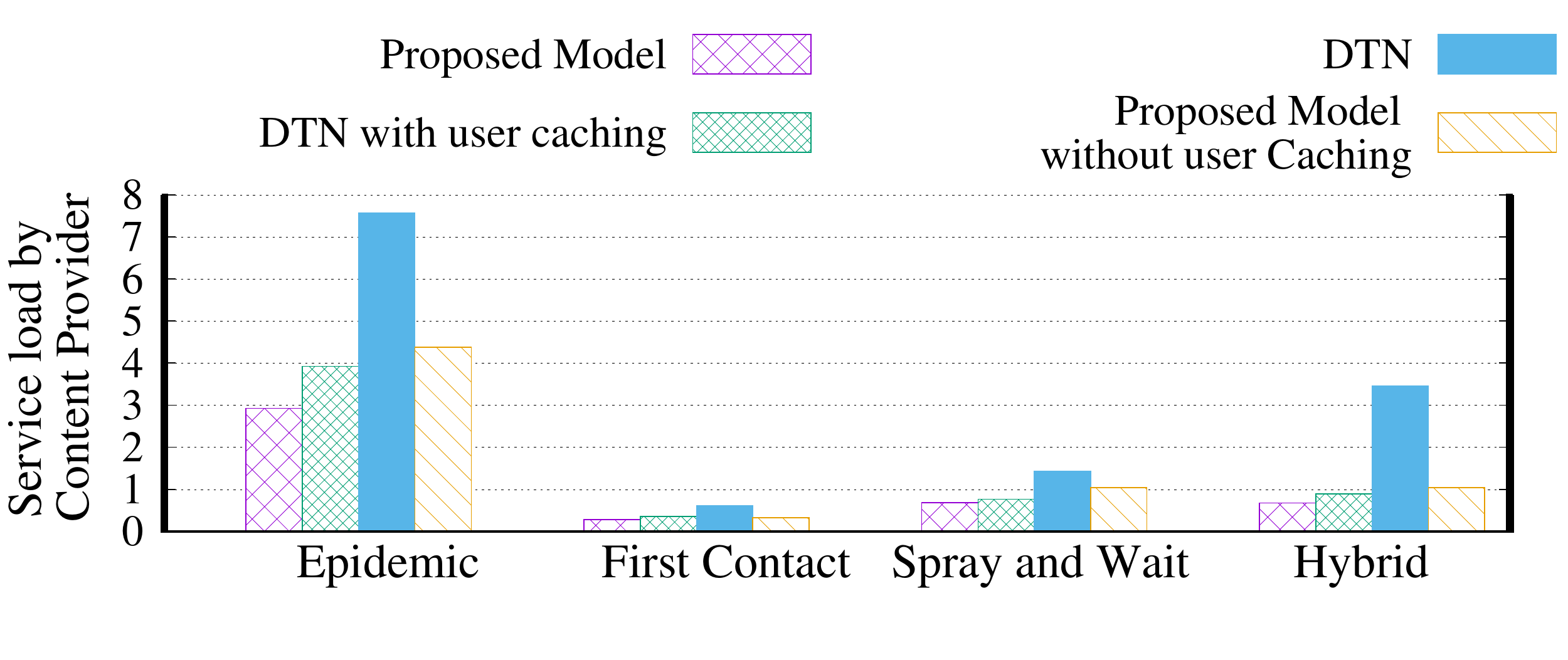}
	\caption{Load in the content provider in the case of contents with popularity that follows the uniform distribution}
	\label{fig:loadUni}
\end{subfigure}
\begin{subfigure}{\columnwidth}
	\includegraphics[height = \columnwidth,angle=270]{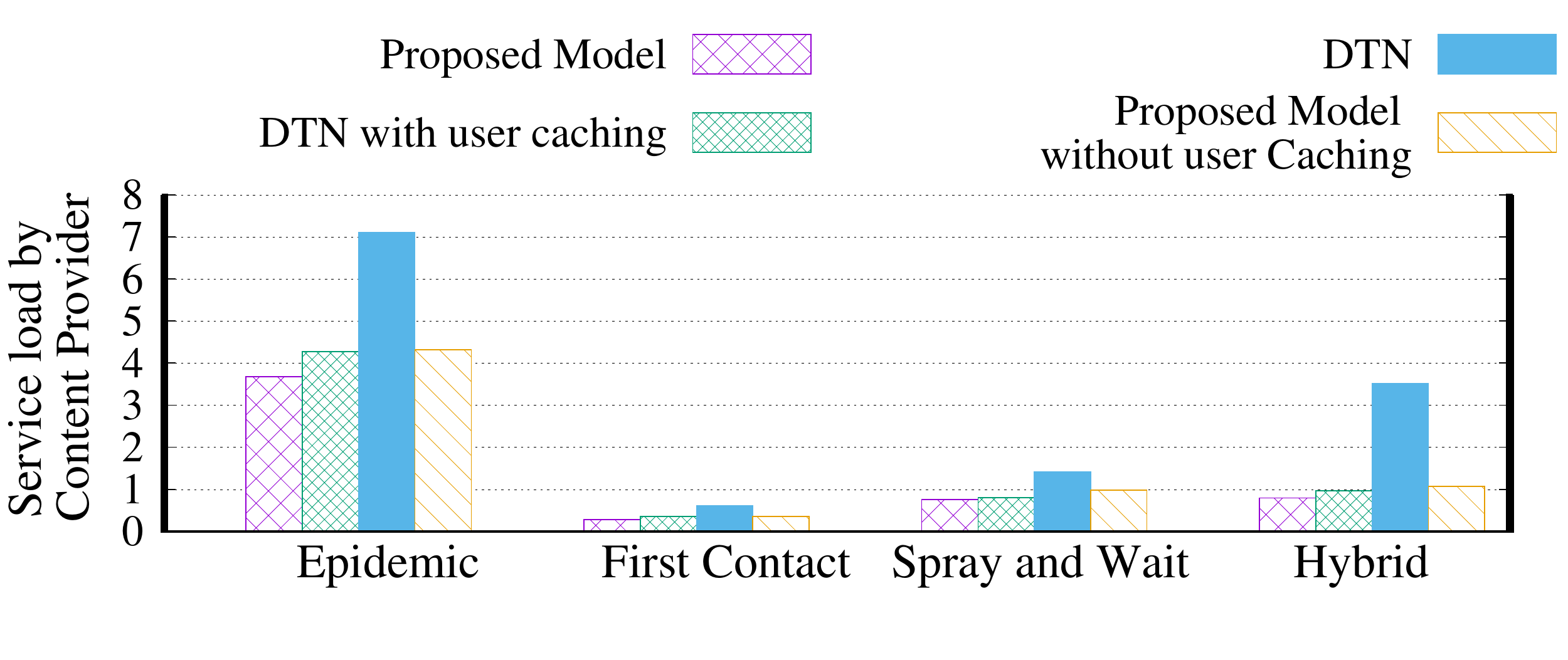}
	\caption{Load in the content provider in the case of contents with popularity that follows the Zipf distribution}
	\label{fig:loadZipf}
\end{subfigure}
\caption{Comparison of our proposal with other frameworks that do not support either CCN functionalities or caching.}
\label{fig:load}
\end{figure}

\subsection{Performance boost of Proposed Architecture}
We measure the performance of our proposal using the metrics that are listed in Table \ref{tab:metrics}. Figure \ref{fig:trafficSplit} shows how the content requests are served of each type of DTN routing protocol. Practically, we show how the hit rates of each content provider type are related. Upon every request, the proposed mechanism uses all the possible ways in parallel in order to download the content as soon as possible. As we can see from both Figure \ref{fig:trafficSplitUn} and Figure \ref{fig:trafficSplitZipf} the content caches in the CAPs together with the caches in the mobile nodes can handle more than 90\% of the requests. Only in the case of the Spray and Wait routing protocol the requests are served by the content producer around 25\% when the content popularity follows the uniform distribution and 20\% when they follow the Zipf distribution.

In order to measure the contribution of the CCN mechanisms and the caches in the mobile users and in the CAPs, we implemented three simpler mechanisms and we compare them with our proposal in Figure \ref{fig:load}. The first one is a simple content search using the DTN mechanism, i.e., that is operating as a request-response application on top of the DTN routing protocols and is denoted by \textit{DTN}. The second one is an improved version of the first one that has content caches. Each content cache can store 10 objects. This mechanism is denoted by \textit{DTN with user caching}. The last one is the same as our proposal but without caching in the mobile users and is denoted by \textit{Proposed Model without user Caching}. As we can see from both \ref{fig:loadUni} and \ref{fig:loadZipf}, our model is under-loading the content providers more than the other competitors. It is worth mentioning that in the case of Epidemic routing, the content provider is overloaded because unlimited number copies of each request is generated until the request reaches the content provider.

\begin{figure}[t]
\centering
	\includegraphics[height = \columnwidth,angle=270]{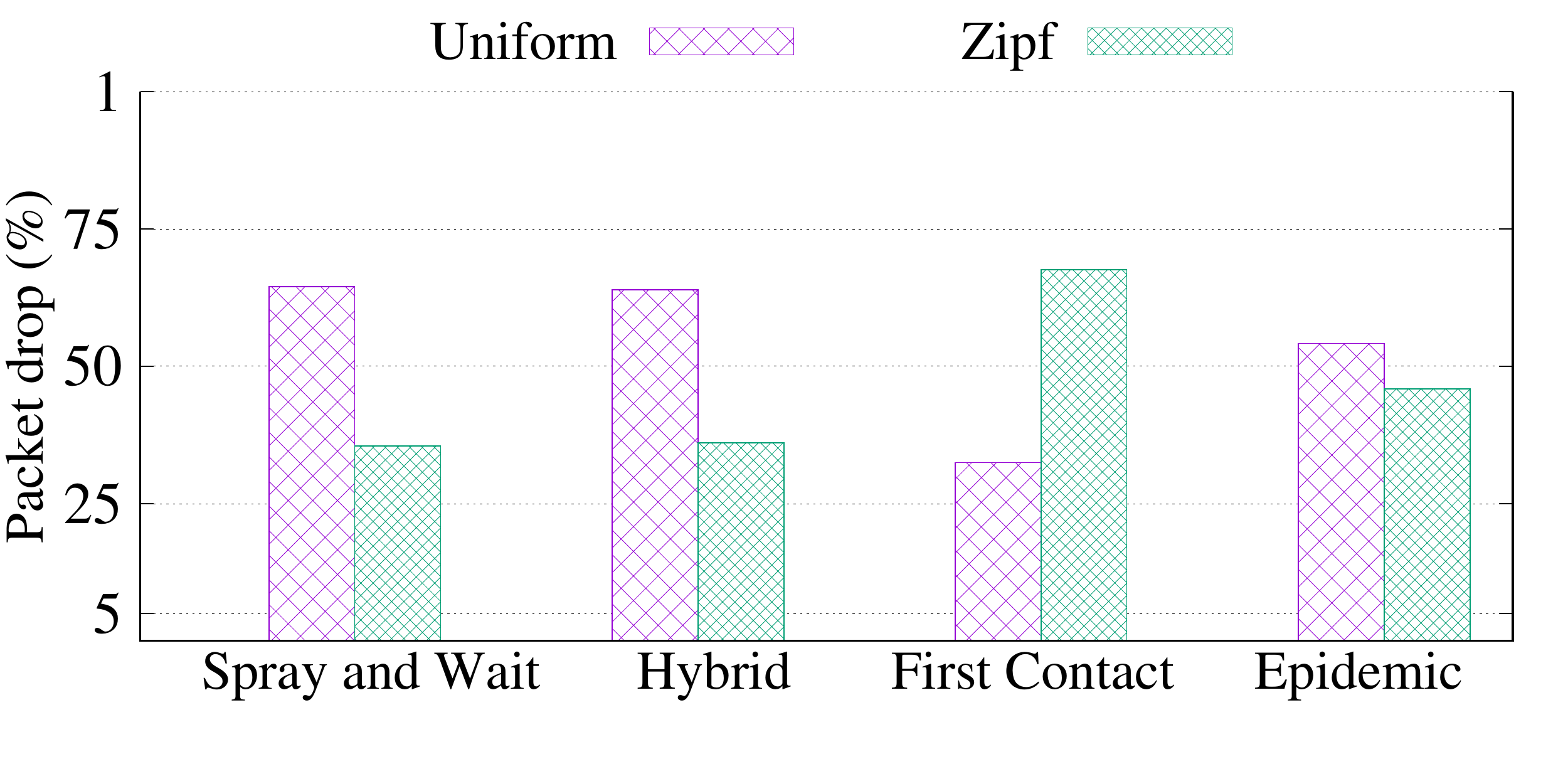}
	\caption{The benefit of CCN in the DTN routing protocols in terms of packet drop. \vspace{-0.3cm}}
	\label{fig:packetDrop}
\end{figure}

Next, we present in Figure \ref{fig:packetDrop} the benefit of using CCN mechanisms in conjunction with the DTN routing protocols because they filter the requests and stop forwarding identical packets. We observe that if user caching is not used, the number of duplicate packets significantly increase. This is happening because the request packets stay in the network longer to reach the potential content provider, and, hence the communication overhead in terms of additional traffic (interest/data packet) increases. Our proposal detects those duplicates and drops them accordingly. For instance, in the case of uniform distribution and multi-copy routing (e.g., Epidemic, Spray and Wait and Hybrid routing), we observe that more than 50\% of the duplicate packets are reduced in our proposal as compared to our proposed model without user cache. This is happening because these protocols are producing multiple copies per request and each content has the same chance of being requested multiple times and being found in a nearby user's pending interest table. On the other hand, in the case of First contact (single copy routing), we observe 27\% duplicate packet reduction. In the case of Zipf content popularity and First Contact, we observe more than 60\% duplicate packets reduction. This is because without the user cache, the requests take a longer time to reach the potential content provider, whereas the other three DTN protocols can potentially reach the content provider faster than First Contact.  

\begin{figure}[t]
\centering
	\includegraphics[height = \columnwidth,angle=270]{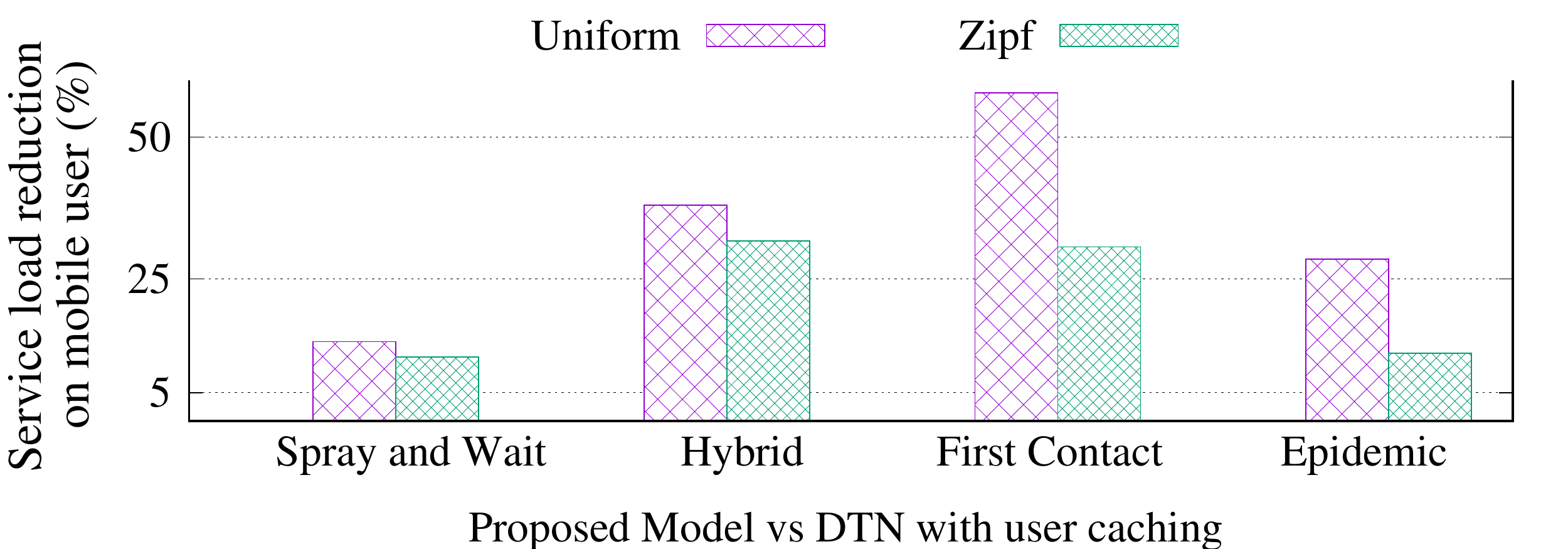}
	\caption{Our proposal vs DTN with user caching but without CCN functionalities. \vspace{-0.3cm}}
	\label{fig:userload}
\end{figure}

\begin{figure}[t]
\centering
\begin{subfigure}{\columnwidth}
	\includegraphics[height = \columnwidth,angle=270]{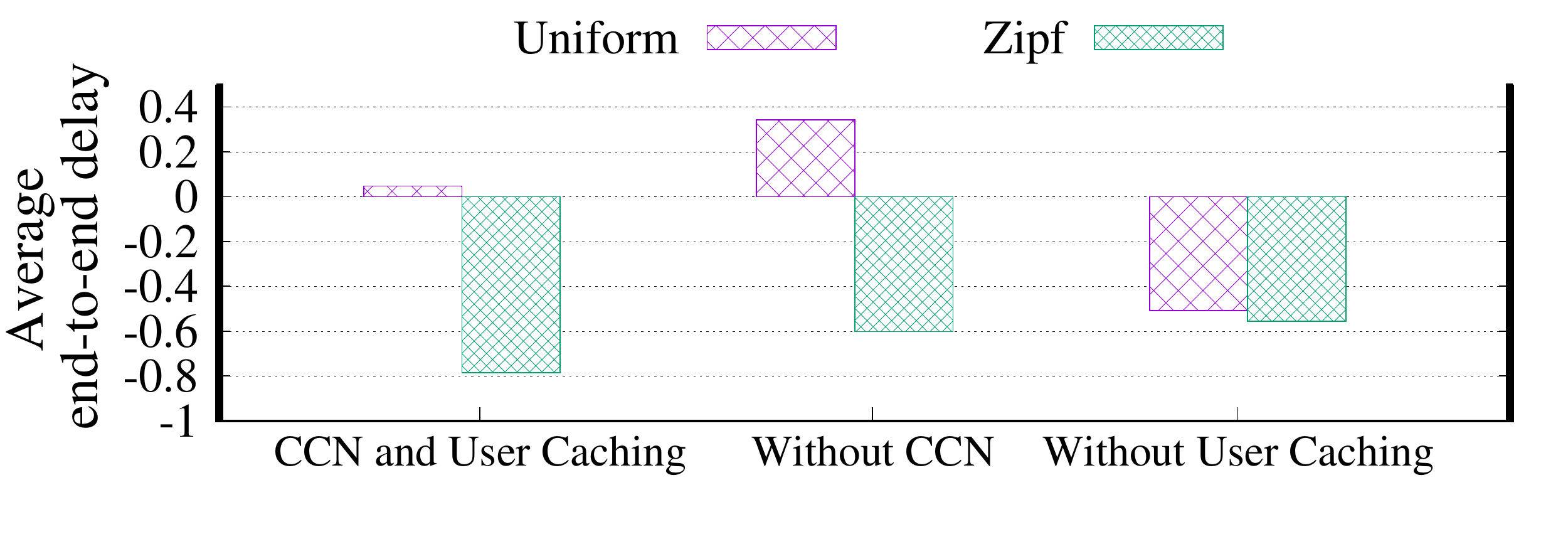}
	\caption{Epidemic routing}
	\label{fig:e2edelayEP}
\end{subfigure}
\begin{subfigure}{\columnwidth}
	\includegraphics[height = \columnwidth,angle=270]{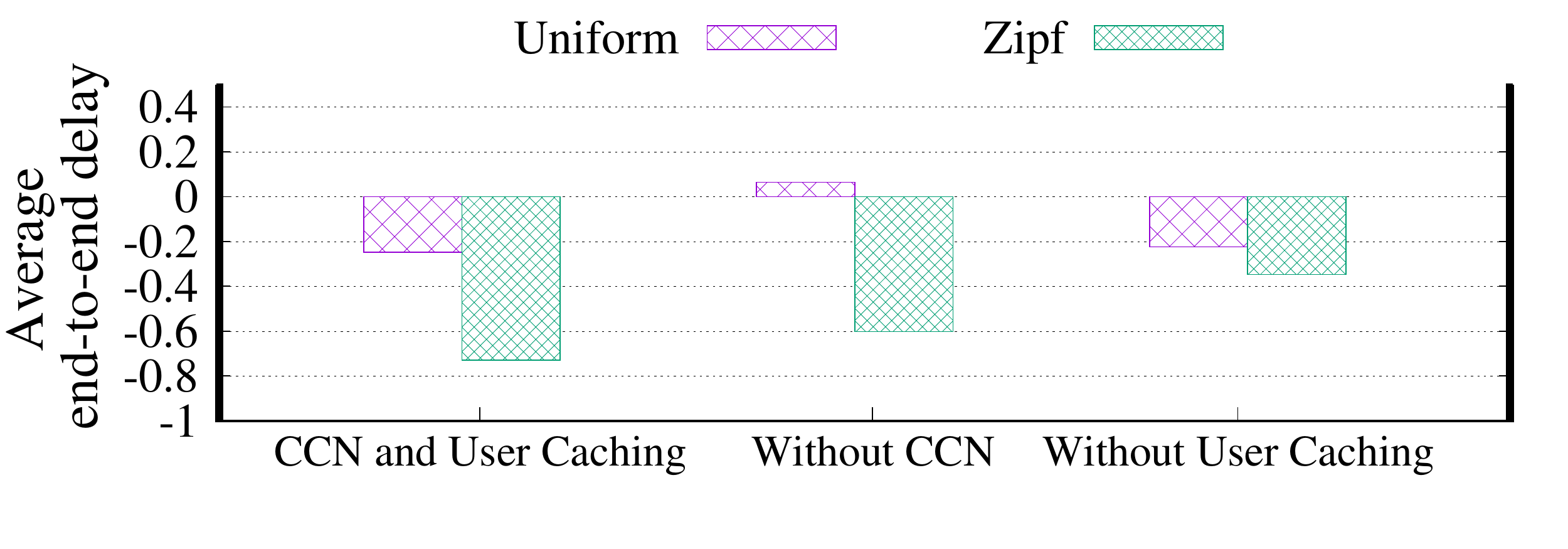}
	\caption{First Contact}
	\label{fig:e2edelayFC}
\end{subfigure}
\begin{subfigure}{\columnwidth}
	\includegraphics[height = \columnwidth,angle=270]{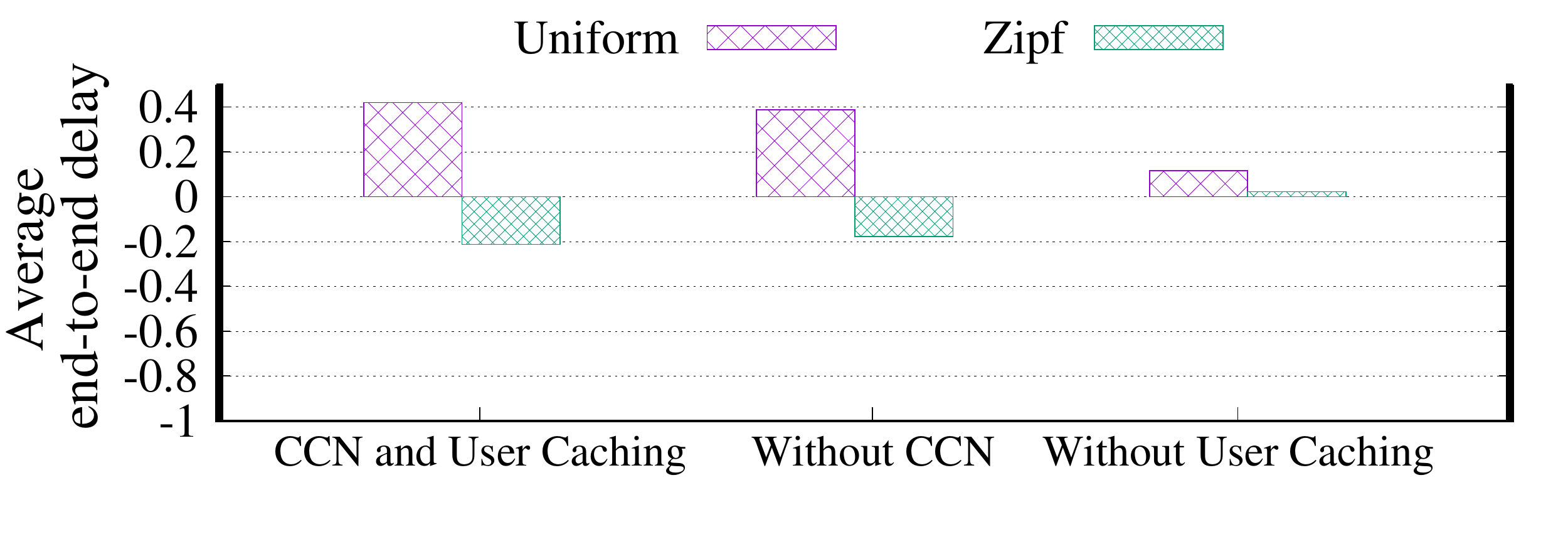}
	\caption{Spray and Wait}
	\label{fig:e2edelaySnW}
\end{subfigure}
\begin{subfigure}{\columnwidth}
	\includegraphics[height = \columnwidth,angle=270]{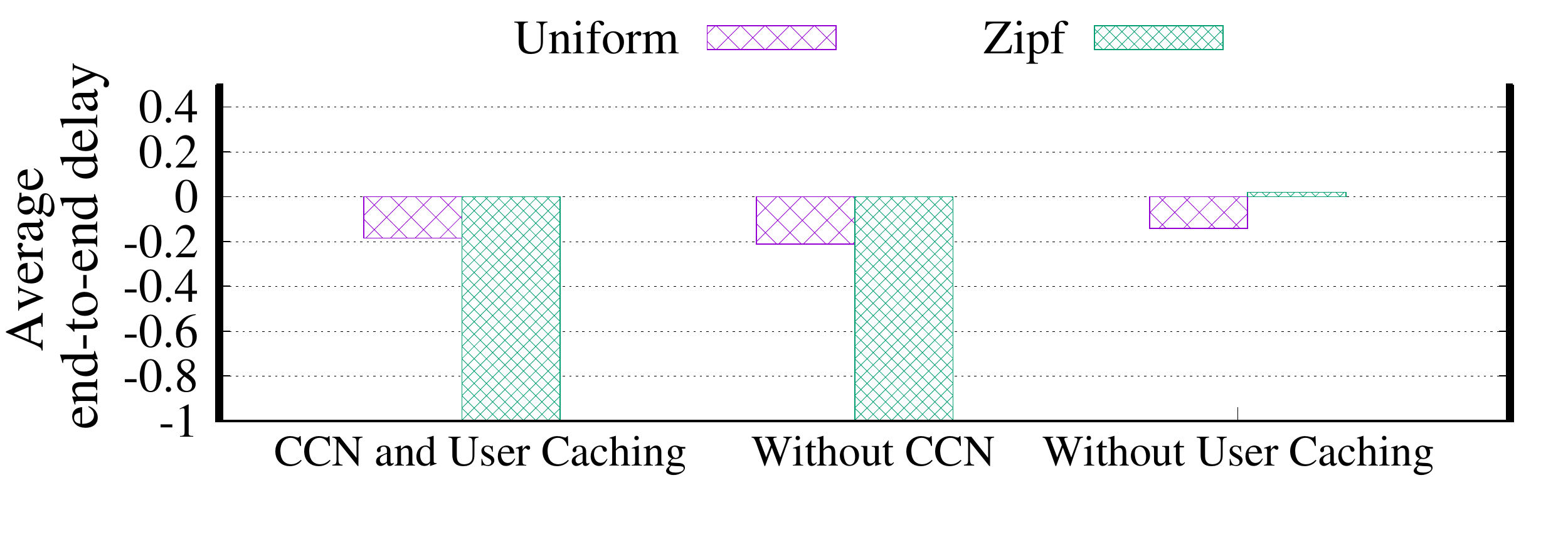}
	\caption{Hybrid}
	\label{fig:e2edelayH}
\end{subfigure}
\caption{Average end-to-end delay change compared to the mechanism that does not have CCN and caching functionalities.\vspace{-0.3cm} }
\label{fig:e2edelay}
\end{figure}

We also examine the service load on mobile users with our proposal to compare \textit{DTN with user caching}. Figure~\ref{fig:userload} shows that our model reduces the service load on the mobile user in all routing. Especially in First Contact routing, the service load on the mobile user is significantly reduced by 57$\%$ when the content popularity follows uniform distribution. On the other hand, when Hybrid and Epidemic routing is used, the service load is reduced by 37$\%$ and 28$\%$ respectively. This is because First contact generates single copy for each request, whereas others use multiple copies. Multiple copies increase the probability of reaching the content provider faster. Service load is not significantly reduced (10$\%$) by the Spray and Wait routing due to a limited number of message copies. We observe that in the case of Zipf content popularity, service load reduction (approximately 10$\%$) on mobile users by Spray and Wait routing is almost similar to Epidemic routing.

Furthermore, we examine the changes in the average delay of the content retrieval in Figure \ref{fig:e2edelay}. As expected, we had a decrease in the delay in most of the cases because of the caching mechanisms. Especially in the case of contents with popularity that follows the Zipf distribution, the contents were accessed faster because they were cached somewhere nearby. However, there are cases where the delay can be increased because there are not many requests for contents in the Spray and Wait routing protocol with contents that follow the uniform distribution (Figure \ref{fig:e2edelaySnW}). 

\section{Conclusion and Future Work}\label{sec:concl}

In this paper, we investigated the possibility of using mobile users in improving the performance of content delivery. For this, we explain the necessary required modifications in the conventional CCN mechanism in order for it to be functional in a DTN environment. Furthermore, we present a mathematical model of the content centric networking framework that exploits the opportunistic communications among mobile users. The proposed framework is implemented in ONE simulator to evaluate the concept. The simulation result shows that caching on mobile devices and cellular access points can improve the content retrieval time by more than 50$\%$, while the proportion of the requests that are delivered from other mobile devices can be more than 75$\%$ in many cases. Our next steps will be focused on the development of caching policies and on various types of contents that are application dependent. Moreover, we plan to consider incentives that motivate mobile users to cooperate and store other content.

\bibliographystyle{IEEEtran}
\bibliography{biblio}

\begin{IEEEbiography}[{\includegraphics[width=1in,height=1.25in,clip,keepaspectratio]{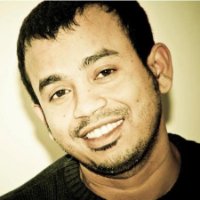}}]{Hasan M A Islam}
received his Bachelor degree in Computer Science and Engineering from Bangladesh University of Engg and Techology which is the top ranked university in Bangladesh in 2008. In 2013, he received his M.Sc degree in Networking and Services from University of Helsinki. He is currently working as a Doctoral Candidate in the Department of Computer Science, Aalto university. His research interests include Information Centric Networking, Communication Network Architecture, and Network protocols.
\end{IEEEbiography}

\vspace{-2.2cm}

\begin{IEEEbiography}[{\includegraphics[width=1in,height=1.25in,clip,keepaspectratio]{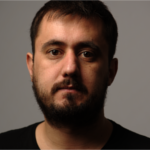}}]{Dimitris Chatzopoulos}
received his Diploma and his MSc in Computer Engineering and Communications
from the Department of Electrical and Computer Engineering of University of Thessaly, Volos, Greece. He is currently a PhD student at the Department of Computer Science and Engineering of The Hong Kong University of Science and Technology and a member of Symlab. His main research interests are in the areas of device-to-€"device ecosystems, mobile computing, mobile augmented reality and cryptocurrencies.
\end{IEEEbiography}
\vspace{-2.2cm}

\begin{IEEEbiography}[{\includegraphics[width=1in,height=1.25in,clip,keepaspectratio]{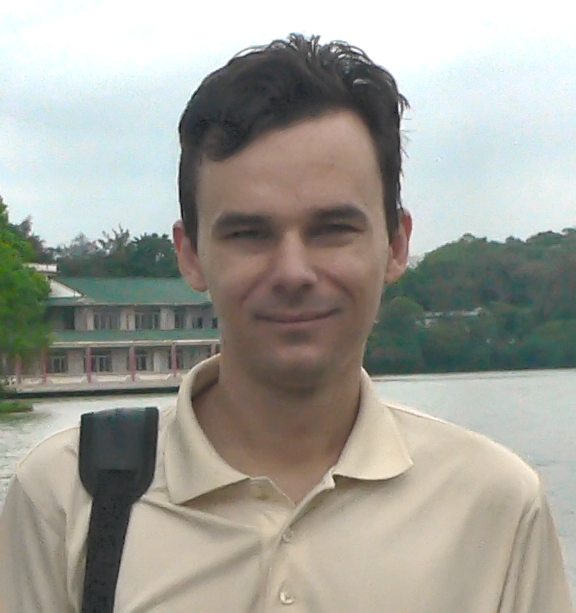}}]{Dmitrij Lagutin} received his M.Sc (Tech) degree in 2005 from 
Helsinki University of Technology and a D.Sc. (Tech) degree in 2010 from 
Aalto University, Finland. He is currently working as a project manager 
and postdoctoral researcher in the EU Horizon 2020 POINT project. 
Previously he worked as a researcher in several research projects at 
Helsinki University of Technology and Aalto University, including EU FP7 
PSIRP and PURSUIT projects. His research interests include network 
security and privacy, future network technologies, and the Internet of 
Things.
\end{IEEEbiography}

\vspace{-2.2cm}

\begin{IEEEbiography}[{\includegraphics[width=1in,height=1.25in,clip,keepaspectratio]{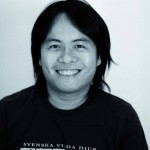}}]{Pan Hui}
received his Ph.D degree from the Computer Laboratory, University of Cambridge, and earned both his MPhil and BEng from the Department of Electrical and Electronic Engineering, University of Hong Kong. He is currently a faculty member of the Department of Computer Science and Engineering at the Hong Kong University of Science and Technology where he directs the HKUST-DT System and Media Lab. He also serves as a Distinguished Scientist of Telekom Innovation Laboratories (Tlabs) Germany and an adjunct Professor of social computing and networking at Aalto University Finland. Before returning to Hong Kong, he spent several years in T-labs and Intel Research Cambridge. He has published more than 150 research papers and has some granted and pending European patents. He has founded and chaired several IEEE/ACM conferences/workshops, and has been serving on the organising and technical program committee of numerous international conferences and workshops including ACM SIGCOMM, IEEE Infocom, ICNP, SECON,
MASS, Globecom, WCNC, ITC, ICWSM and WWW. He is an associate editor for IEEE Transactions on Mobile Computing and IEEE Transactions on Cloud Computing, and an ACM Distinguished Scientist.
\end{IEEEbiography}

\vspace{-2.2cm}

\begin{IEEEbiography}[{\includegraphics[width=1in,height=1.25in,clip,keepaspectratio]{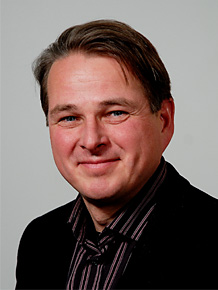}}]{Antti Yl{\"a}-J{\"a}{\"a}ski}
 is the vice head of the Department Computer Science, Aalto University. Prior to his current position, he was with Nokia 1994-2004 in several research and research management positions, with focus on future Internet, mobile networks, applications, services and service architectures. He has supervised 231 masters thesis and 21 doctoral dissertations during his professorship in Aalto University 2002-2015; he has published over 100 peer-reviewed articles and he holds several international patents. His current research interest is focused on mobile and cloud computing, crowdsourcing and energy efficient computing and communications.
\end{IEEEbiography}

\end{document}